\documentclass[aip,cha,amsmath,amssymb,reprint,author-numerical]{revtex4-1}

\usepackage{graphicx}% Include figure files
\usepackage{dcolumn}% Align table columns on decimal point
\usepackage{bm}% bold math
\usepackage{hyperref}
\usepackage{color}

\begin{document}

\preprint{AIP/123-QED}

\title{Chimeras in digital phase-locked loops} %Title of paper


\author{Bishwajit Paul}
\affiliation{Chaos and Complex Systems Research Laboratory, Department of Physics, University of Burdwan, Burdwan 713 104, West Bengal, India.}
\author{Tanmoy Banerjee}
\email{tbanerjee@phys.buruniv.ac.in}
\affiliation{Chaos and Complex Systems Research Laboratory, Department of Physics, University of Burdwan, Burdwan 713 104, West Bengal, India.}

\received{:to be included by reviewer}
\date{\today}

\begin{abstract}
Digital phase-locked loops (DPLLs) are nonlinear feedback-controlled systems that are widely used in electronic communication and signal processing applications. In most of the applications they work in coupled mode, however, vast of the studies on DPLLs concentrate on the dynamics of a single isolated unit. In this paper we consider both one- and two-dimensional networks of DPLLs connected through a practically realistic nonlocal coupling and explore their collective dynamics. For the one-dimensional network we analytically derive the parametric zone of stable phase-locked state in which DPLLs essentially work in their normal mode of operation. We demonstrate that apart from the stable phase-locked state, a variety of spatiotemporal structures including chimeras arise in a broad parameter zone. For the two-dimensional network under nonlocal coupling we identify several variants of chimera patterns, such as strip and spot chimeras. We identify and characterize the chimera patterns through suitable measures like local curvature and correlation function. Our study reveals the existence of chimeras in a widely used engineering system, therefore, we believe that these chimera patterns can be observed in experiments as well. 
\end{abstract}

\pacs{05.45.Xt, 05.45.Gg, 05.45.Pq}% insert suggested PACS numbers in braces on next line
\keywords{digital phase-locked loop, non-local coupling, chimera, spot chimera, 2D-network.}

\maketitle 

\begin{quotation}
The phase locking is an intriguing phenomenon observed in the field of physics, biology and engineering. Based on the technique of phase locking de Bellescize invented the phase-locked loop (PLL) in 1932, which has been in the heart of the electronic communication system and signal processing. Starting from the analog television receivers in the late thirties of the previous century, PLLs are now a days used in applications, such as hard-disk drives, modems and atomic force microscopy. With the advent of digital IC technology, PLLs are rapidly replaced by their digital counterpart, namely the  digital phase-locked loops (DPLLs). In the initial years, a single unit of DPLL was sufficient for a particular application. However, modern communication systems demand networking and parallel processing, and therefore, DPLLs typically have to work in cascade. Being a nonlinear system, a vast amount of studies have been dedicated to understand the dynamics of a single DPLL, however, only a few studies are available on networks of coupled DPLLs. In general, the dynamics of coupled nonlinear systems are complex and challenging to understand, especially with the discovery of new spatiotemporal patterns, like chimera states.  

This paper deals with the coupled dynamics of DPLLs and reports the occurrence of chimeras in one- and two-dimensional networks of DPLLs under the practical coupling scheme. We show that apart from the stable phase-locked state, which is the desired mode of operation of DPLLs, the networks show several chimera patterns in a broad parameter space. Our study is significant in the sense that it established the presence of chimeras in a widely used engineering system that can be implemented in hardware.  
\end{quotation}

%**************************(Introduction)***********************
\section{Introduction}
\label{sec:intro}
The phenomenon of phase locking has been fascinating the researchers in the field of physics, biology and engineering \cite{Pikovsky_2001,sync}. In the field of engineering, phase-locked loops (PLLs) are important units of electronic communication and signal processing systems, whose working principle is essentially based on phase locking techniques~\cite{gardner79,best}. With the advent of digital IC technology, PLLs are rapidly replaced by digital phase-locked loops (DPLLs) due to several advantages of the latter~\cite{lin81}. DPLLs are nonlinear discrete feedback-controlled systems that have wide applications in integrated electronic devices and coherent communication systems \cite{gardner79,lin81,leo14}. They are used in the receiver segment of a communication system as data-clock recovery circuit and frequency demodulators \cite{dpll_book}. Also, DPLL is an important building block of various electronic systems such as digital signal processors and hard-disk drives~\cite {zol01}. Apart from their application potentials, DPLLs (and PLLs) exhibit rich complex phenomena such as bifurcations and chaos \cite{banerjee2008,leo09}, which have widely been studied in the literature.

In modern applications, typically DPLLs work in cascade, i.e., they operate under coupled condition. 
%The coupling scheme varies with the field of application: 
Take for instance the problem of clock skew in clock synchronizers: in a multiprocessor system under the application of parallel processing where the units are spatially separated the propagation time of clock signal to reach different processors are different, this is called the clock skew~\cite{leo05}. In order to avoid clock skew, in these systems it is essential to use DPLLs (or PLLs) in coupled mode to synchronize the units. Therefore, it is absolutely important to understand how an array of coupled DPLLs behave. However, surprisingly, in the literature only a few studies deal with the collective behavior of coupled DPLLs: The temporal behavior of two coupled DPLLs was studied in Ref.~\onlinecite{bern1,*bern2}; \citet{stog} treated coupled DPLLs as pulse coupled oscillators and derived explicit formulas for the transient time to lock, stability of the synchronized state, and the period of the bifurcating solution at the onset of instability. The collective dynamics of a network of DPLLs with {\it nearest neighbor coupling} topology has recently been reported by the present authors \cite{T_Banerjee_2014,B_Paul_2016}: it was shown that the network depicts the formation of a variety of spatial and spatiotemporal patterns, like frozen random pattern, pattern selection, spatiotemporal intermittency and spatiotemporal chaos. In an another context, \citet{dan2} studied Boolean phase oscillators, which are essentially the core of all-digital phase-locked loops (ADPLLs), and experimentally explored and characterized their coupled behaviors. Later on, the study was extended for a larger network (one-dimensional) and significantly complex spatiotemporal patterns (including chimera~\cite{Y_Kuramoto_2002}) were observed in an experimental set up~\cite{dan1,dan3}. However, contrary to DPLLs, which are modeled as discrete time maps, in Refs.~\onlinecite{dan1,dan2,dan3} Boolean network was modeled as continuous time phase oscillators.
%and also to realize variable coupling strength in a Boolean network state dependent time delay was considered. 

In a real network of spatially separated DPLLs local coupling is weaker to achieve synchronization by overcoming the disorder caused by the local dynamics. On the other hand, although global coupling is conducive for synchronization, however, this form of coupling is massive and costly to implement in a real network. In this context, the nonlocal coupling topology is the most general one: depending upon the target dynamics one can make a trade-off between the coupling range and the coupling strength. 
Apart from coupling topology, dimensionality of a network also plays a crucial role in determining the overall dynamics. In comparison with the one-dimensional network, a two-dimensional network is much more general yet relatively less explored in the context of coupled oscillators. In the case of a DPLL based network also, a two-dimensional network is a much realistic network structure, however, hitherto its dynamics has not been studied.  
%The proposed two dimensional network of DPLLs is more real and physical compared to the previously studied networks, which were only the mathematical model networks of some oscillators \cite{Chad_Gu_2013, Schmidt_2017, Tian2017}.

Unlike local coupling \cite{T_Banerjee_2014,B_Paul_2016}, it is expected that a nonlocal coupling can give rise to much more complex spatiotemporal behaviors, such as chimera patterns, which is in the center of recent research.
The chimera state, discovered by Kuramoto and Bottogtokh \cite{Y_Kuramoto_2002}, is defined as the hybrid dynamical spatiotemporal state of a network in which different subsets of  identical oscillators spontaneously divided into two groups, namely, synchronized (coherent) and desynchronized (incoherent) groups. In the initial years most of the studies  explored several aspects of chimera states using theoretical and numerical analysis (see two recent reviews on chimera in Ref.~\onlinecite{chireview} and Ref.~\onlinecite{schoell_rev}).  
Later on, chimera states were also observed experimentally: The first experimental observation of chimera was reported in the optical coupled map lattice system~\cite{raj} and electrochemical system~\cite{M_R_Tinsley_2012}. Chimeras have later been observed experimentally in optoelectronic systems \cite{Larger_2015}, mechanical systems \cite{Kapitaniak_2014,Erik_2013}, chemical oscillators \cite{nennatphys_18}, and Boolean networks~\cite{dan1}. 
The current interest on the chimera state can be attributed to its possible connection with the unihemispheric slow-wave sleep of some migratory birds and aquatic mammals \cite{Rattenborg_2000, Rattenborg_2006}.
Several real world systems also exhibit chimera state, e.g., social networks \cite{Avella_2014}, brain network \cite{epileptic}, ecological networks \cite{banps,banlr}, quantum systems \cite{schoell_qm} and SQUID metamaterials \cite{hiz16} to name a few. Recent studies also reveal many interesting chimera patterns in two-dimensional~\cite{Schmidt_2017,Tian2017} and three-dimensional networks \cite{3d2018}.
%For recent reviews on the chimera states see Ref.~\onlinecite{chireview} and Ref.~\onlinecite{schoell_rev}.

In most of the experimental studies on chimeras the main goal was to observe chimera patterns in man made systems under certain coupling and initial conditions: however, bulk of the systems under study (with a few exceptions) are of academic interest. Therefore, it is a natural question to ask, whether a real engineering system with wide application potentiality, like DPLL, can show chimera state under its {\it normal} coupling condition.

Motivated by the above discussion, in this paper we study the dynamics of nonlocally coupled DPLLs in one- and two-dimensional networks. In both the cases depending upon coupling and local dynamics we show that several chimera patterns emerge. For the 1D case we analytically obtain the condition for the phase-locked state, which is the normal and desirable mode of operation for DPLLs. In the case of two-dimensional (2D) network we observe several interesting 2D chimera patterns, such as spot and strip chimeras. We characterize the chimera states by using quantitative measures such as the local curvature and local correlation function. Finally we also discuss the importance of our study in the context of engineering.

%%%%%%%%%%%%%%%SECTION1:ISOLATED DPLL%%%%%%%%%%%%%%%%%%%%%%%%%%%%%%%%
\section{System description and mathematical model}

\subsection{Isolated ZC1-DPLL}
\begin{figure}
\begin{center}
\includegraphics[width=.49\textwidth]{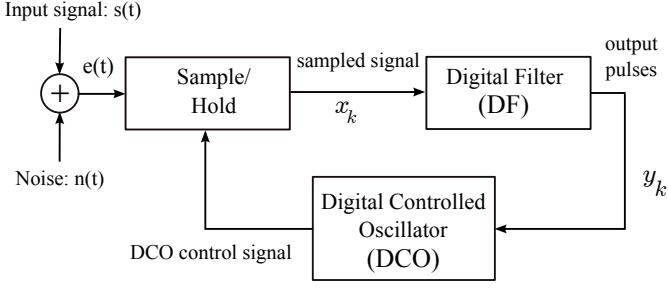}
\caption[Isolated ZC1-DPLL]{The figure shows the block diagram of an isolated ZC1-DPLL.}
    \label{isolated_zc1dpll_4}
\end{center}
\end{figure}

The functional block diagram of a ZC1-DPLL is shown in the Fig.~\ref{isolated_zc1dpll_4}. It consists of a sample and hold block, a digital filter (DF) and a digital-controlled oscillator (DCO). Let us assume that a noise free $(n(t)=0)$ sinusoidal incoming signal $e(t)=A_0\sin(\omega_i t+\theta_0)$ is fed to the sample and hold block ($A_0$: amplitude, $\omega_i$: angular frequency and $\theta_0$: initial phase). This signal is sampled by the sample and hold block at each positive going zero-crossing edges of the pulse signals from the DCO. The sampled signal $x_k=e(t_k)$ at the $k^{th}$ sampling instant $t_k$ is modified by the DF that generates $y_k$ ($k=0,1,2,\cdots$), which in turn controls the DCO to determine the next sampling instants of the sample and hold block. 
The incoming signal $e(t)$ can be rewritten in terms of the free-running angular frequency of the DCO, $\omega_0$, as
\begin{equation}\label{in_sig_ch4}
e(t)=A_0\sin\left(\omega_0t+\theta(t)\right)=A_0\sin(\phi(t)),
\end{equation}
where $\theta(t)=(\omega_i-\omega_0)t+\theta_0$. The instantaneous time period of the DCO at the $k^{th}$ instant is determined by \cite{dpll_book}: $T_k=t_k-t_{k-1}$. The sequence of samples $\{x_k\}$ generate control sequences $\{y_k\}$ that are employed to determine the successive time period by the algorithm \cite{banerjee2005}: $T_{k+1}=T-y_k$, where $T(=2\pi/\omega_0)$ corresponds to the nominal period of the DCO. Without any loss of generality we assume $t_0=0$, then the sampling instants are given by~\cite{lin81}
\begin{equation}\label{eq_tk_ch4}
t_k=kT-\sum_{i=0}^{k-1}y_i.
\end{equation}
The phase error of the incoming signal in Eq.~\ref{in_sig_ch4} at the $k^{th}$ sampling instant is $\phi_k=\phi(t_k)=\omega_0t_k+\theta(t_k)=\omega_0t_k+\theta_k$; then using Eq.~\eqref{eq_tk_ch4} we get
\begin{equation}
\phi_k=\omega_0\left(kT_0-\sum_{i=0}^{k-1}y_i\right)+\theta_k.
\end{equation}
The first term $\omega_0kT_0$ is integral multiple of $2\pi$ and the effective residual phase error becomes
 \begin{equation}\label{phase_err_ch4}
\phi_k=\theta_k-\omega_0\sum_{i=0}^{k-1}y_i.
\end{equation}
For a first-order ZC1-DPLL the digital filter is a zero-th order filter with no memory element having a constant gain $G_0$ (say). Therefore, the control signal is given by $y_k=G_0x_k$. Then from Eq.~\eqref{phase_err_ch4} one can get the phase error equation of the system as
\begin{equation}\label{syst_eq_iso_ch4}
\phi_{k+1}=\Lambda_0+\phi_k-K_1\xi\sin(\phi_k),
\end{equation}
where $\xi=\omega_i/\omega_0$ is the normalized incoming frequency, $\Lambda_0=2\pi(\xi-1)$ and $K_1=A_0\omega_0G_0$. $K_1$ is called the closed loop gain of the ZC1-DPLL. Physically all $\{\phi_k\}$s are modulo $2\pi$ quantity that are bounded in the interval $[-\pi:\pi]$. Eq.~\eqref{syst_eq_iso_ch4} represents the system equation of an isolated ZC1-DPLL.

%%%%%%%%%%%%%%%SECTION2:1D network of DPLLs%%%%%%%%%%%%%%%%%%%%%%%

\subsection{One-dimensional network of ZC1-DPLLs with nonlocal coupling}\label{1dnet}

\begin{figure}[!b]
  \begin{center}
    \leavevmode
    \ifpdf
      \includegraphics[height=3in]{}
    \else
      \includegraphics[width=.5\textwidth]{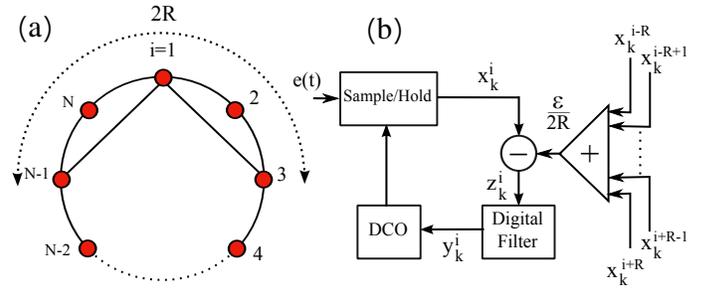}
    \fi
    \caption[Schematic diagram of an one-dimensional network of ZC1-DPLL with nonlocal coupling]{(Color online) The schematic diagram of a 1D network of ZC1-DPLLs with nonlocal coupling. (a) The nonlocal coupling scheme with periodic boundary condition. We  draw only the coupling links of $i=1^{st}$ ZC1-DPLL, which is coupled with $R$ ZC1-DPLLs on either side (here $R=2$). (b) The schematic of the practical circuit realization for the $i^{th}$ ZC1-DPLL in the network. $x^i_k$ is the sampled value for the $i^{th}$ ZC1-DPLL at the $k^{th}$ sampling instant. Here $Z^i_k$ is obtained from the difference between $x^i_k$ and the weighted sum of the outputs of sample and hold blocks of other nodes of the network.}
    \label{fig_system_1d_nonlocal}
  \end{center}
\end{figure}
We consider a one-dimensional (1D) network of nonlocally coupled $N$ identical ZC1-DPLLs arranged in a ring topology with periodic boundary condition. Figure~\ref{fig_system_1d_nonlocal}~(a) shows the schematic representation of the coupling scheme. The practical coupling architecture in which a network of coupled ZC1-DPLLs normally works is shown in Fig.~\ref{fig_system_1d_nonlocal}(b). According to this scheme the $i^{th}$ node is coupled with $R$ nearest neighbors on either side of its position with a normalized coupling strength $\frac{\epsilon}{2R}$. The sampled input signal for the $i^{th}$ ZC1-DPLL is given by $x^{i}_k$, where the spatial position is indexed by the superscript $i$ ($i=1,2,\cdots N$) and the temporal instant is indexed by a subscript $k$. The coupling scheme can be realized by using an weighted adder, that add the sampled signals $x^{i-R}_k, x^{i-R+1}_k. \dots x^{i-1}_k, x^{i+1}_k, \dots x^{i+R-1}_k, x^{i+R}_k$ obtained from the outputs of the sample and hold blocks of neighboring ZC1-DPLLs. Therefore, the input of the digital filter of the $i^{th}$ ZC1-DPLL is given by
\begin{equation}
%Z^i_k &=& x^i_k-\frac{\epsilon}{2R}(x^{i-R}_k, x^{i-R+1}_k. \dots x^{i-1}_k, x^{i+1}_k, \dots x^{i+R-1}_k, x^{i+R}_k) \nonumber\\
Z^i_k=x^i_k-\frac{\epsilon}{2R}\sum_{j=i-R}^{i+R}(x^j_k-x^i_k).
\end{equation}
The second term in the bracket discards the contribution due to the self-feedback. The output of the digital filter is given by $y^i_k=G_0Z^i_k$ (cf. $G_0$ is the constant gain of the digital filter). Using the same method as used for an isolated ZC1-DPLL one gets the following system equation for the network:
\begin{eqnarray}\label{system_eq_nonlocal_1d}
\phi^i_{k+1}&=&\Lambda_0+\phi^i_k-\xi K_1 \sin\phi^i_k\nonumber\\
&&+\:\frac{\epsilon \xi K_1}{2R}\sum^{i+R}_{j=i-R}\left[\sin(\phi^j_{k})-\sin(\phi^i_k)\right].
\end{eqnarray}
Interestingly, Eq.~\eqref{system_eq_nonlocal_1d} takes the standard form of a coupled map lattice (CML) system~\cite{kan1} with nonlocal coupling.
%Since $R$ represents the range of nonlocal coupling, therefore, for $R=0$, the network is an ensemble of $N$ isolated ZC1-DPLLs, whereas $R=1$ corresponds to the nearest neighbor coupling. On the other hand if total number of sites becomes odd, then for $R=\frac{N-1}{2}$ the feedback term is contributed by all the rest $N-1$ sites in the lattice and this corresponds to global coupling. But if the total number of sites are even, then the symmetrical coupling up to the entire lattice (global coupling) is not perfectly defined. Thus for $1<R<\frac{N-1}{2}$ the feedback is neither local nor global and this case corresponds to nonlocal coupling.

%%%%%%%%%%%%%%SECTION3:SYSTEM DESCRIPTION%%%%%%%%%%%%%%%%%%%%%%%%%%%%
\section{Stability analysis}
\subsection{Isolated ZC1-DPLL}
The steady stare phase-error of the isolated system \eqref{syst_eq_iso_ch4} is given by: $\phi_s=\sin^{-1}\left(\frac{\Lambda_0}{\xi K_1}\right)$. Using $|f^\prime(\phi_s)|<1$, where $f(\phi)=\Lambda_0+\phi-K_1\xi\sin(\phi)$ and $f^\prime(\phi_s)=1-\xi K_1\cos(\phi_s)$ one gets the following condition for the stable phase-locked state \cite{Osborne}: $0<(K_1\xi)^2-\Lambda_0^2<4$. For the normal operation of a DPLL this condition has to be satisfied.

\begin{figure}[!htbp]
  \begin{center}
    \leavevmode
    \ifpdf
      \includegraphics[height=3in]{}
    \else
      \includegraphics[width=.48\textwidth]{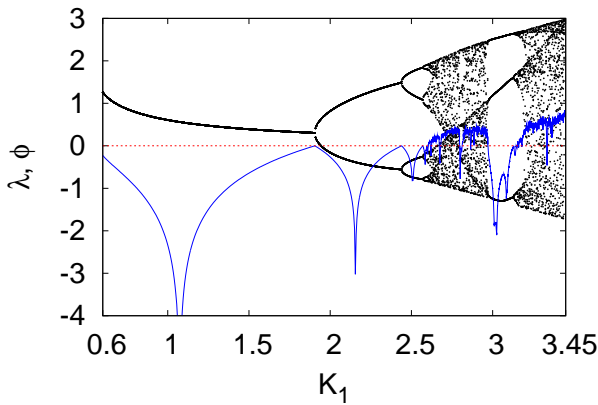}
    \fi
    \caption[Bifurcation diagram of an isolated DPLL]{(Color online) Bifurcation diagram (black points) along with the Lyapunov exponent spectrum (blue line) of an isolated ZC1-DPLL ($\xi=1.1$).}
    \label{isolated_bifur}
  \end{center}
\end{figure}

In the context of ZC1-DPLLs, the closed loop gain, $K_1$, is the most important parameter: larger $K_1$ provides larger frequency acquisition range, broader tracking range and lesser steady state phase errors \cite{lin81}. However, with increasing $K_1$ the system enters into chaotic state through a period doubling cascade (for a detailed dynamical behavior of the system see Ref.~\onlinecite{T_Banerjee_2012}). The bifurcation diagram along with the corresponding Lyapunov exponent spectrum for an isolated ZC1-DPLL are shown in Fig.~\ref{isolated_bifur} for $\xi=1.1$ (the Lyapunov exponent is given by: $\lambda=\lim_{n\rightarrow\infty}\frac{1}{n}\sum_{i=1}^{n}\ln|1-\xi K_1 \cos(\phi_i)|$). 
%Also, the two parameter Lyapunov exponent spectrum in $K_1-\epsilon$ space  shows the whole scenario of the dynamical features of a ZC1-DPLL\cite{T_Banerjee_2012}.
%%%%%%%%%%%%%%SECTION3:STABILITY ANALYSIS%%%%%%%%%%%%%%%%%%%%%%%%%%%%
\subsection[Stability analysis: 1D network of ZC1-DPLLs with non-local coupling]{Stability analysis: one-dimensional network of ZC1-DPLLs with nonlocal coupling}
Here we will derive the condition for which all the ZC1-DPLLs in the network (given by \eqref{system_eq_nonlocal_1d}) achieve a synchronized stable phase-locked state.
The system equation \eqref{system_eq_nonlocal_1d} is equivalent to a set of $N$ difference equations, which can be written as: $\phi^i_{k+1}=J\phi^i_k$, where $J$ is the Jacobian matrix of the system. One can write the Jacobian matrix as
\begin{equation}
J=\left[
\begin{matrix}\label{system_jacobi}
\frac{\partial\phi^1_{k+1}}{\partial\phi^1_k}&\frac{\partial\phi^1_{k+1}}{\partial\phi^2_k}&\frac{\partial\phi^1_{k+1}}{\partial\phi^3_k}&\hdots&\frac{\partial\phi^1_{k+1}}{\partial\phi^{N-1}_k}&\frac{\partial\phi^1_{k+1}}{\partial\phi^N_k}\\
\frac{\partial\phi^2_{k+1}}{\partial\phi^1_k}&\frac{\partial\phi^2_{k+1}}{\partial\phi^2_k}&\frac{\partial\phi^2_{k+1}}{\partial\phi^3_k}&\hdots&\frac{\partial\phi^2_{k+1}}{\partial\phi^{N-1}_k}&\frac{\partial\phi^2_{k+1}}{\partial\phi^N_k}\\
\frac{\partial\phi^3_{k+1}}{\partial\phi^1_k}&\frac{\partial\phi^3_{k+1}}{\partial\phi^2_k}&\frac{\partial\phi^3_{k+1}}{\partial\phi^3_k}&\hdots&\frac{\partial\phi^3_{k+1}}{\partial\phi^{N-1}_k}&\frac{\partial\phi^3_{k+1}}{\partial\phi^N_k}\\
\vdots&\vdots&\vdots&\vdots&\vdots&\vdots\\
\frac{\partial\phi^N_{k+1}}{\partial\phi^1_k}&\frac{\partial\phi^N_{k+1}}{\partial\phi^2_k}&\frac{\partial\phi^N_{k+1}}{\partial\phi^3_k}&\hdots&\frac{\partial\phi^N_{k+1}}{\partial\phi^{N-1}_k}&\frac{\partial\phi^N_{k+1}}{\partial\phi^N_k}\\
\end{matrix}
\right].
\end{equation}
The fixed point of the coupled system \eqref{system_eq_nonlocal_1d} is given by
\begin{equation}\label{ss_1d_nonlocal_ch4}
\phi_s=\sin^{-1}\left(\frac{\Lambda_0}{K_1\xi(1-\epsilon)}\right).
\end{equation}
The Jacobian matrix \eqref{system_jacobi} of the coupled system at the stable synchronized fixed point can be written as
\begin{equation}
J_s=\left(
\begin{matrix}
a & b & \dots & b & 0 & \dots& 0 & b & \dots & b\\
b & a & b & \dots& b & 0 & \dots & 0 & b &\dots \\
b & b & a & b & \dots& b & 0 & \dots & 0 & b \\
\vdots & \vdots & \vdots & \vdots & \vdots & \vdots & \vdots & \vdots & \vdots & \vdots\\
b & \dots & b & 0 & \dots& 0 & b & \dots & b & a\\
\end{matrix}
\right).
\end{equation}
For the symmetrical coupling, $J_s$ becomes a circulant matrix of order $(N \times N)$. It is noteworthy that the first row of the above matrix $J_s$ has the following symmetry:
\begin{equation}
a \quad \overbrace{b \quad \dots \quad b}^\text{$R$ terms} \quad \overbrace{0 \quad \dots \quad 0}^\text{$N-(2R+1)$ terms} \quad \overbrace{b \quad \dots \quad b}^\text{$R$ terms}, \nonumber
\end{equation}
note that, it contains one $a$ term, $2R$ numbers of $b$ terms and $N-(2R+1)$ terms of $0$ (zeros). The next rows can be obtained by shifting the elements to the right side by one unit in each step in the circulant form, where the diagonal terms are
\begin{equation}
a=1-\xi K_1 \cos(\phi_s),
\end{equation}
and the other non-zero off-diagonal terms are
\begin{equation}
b=\frac{\epsilon \xi K_1}{2R}\cos(\phi_s).
\end{equation}
The stability condition of such dynamical system can be determined by finding the eigenvalues of the Jacobian matrix $J_s$. Now, $J_s$ has $N$ number of eigenvalues, namely $\lambda_1,\lambda_2,\dots,\lambda_N$. For the stable synchronized state, absolute values of all the eigenvalues should be less than unity, i.e., $|\lambda_l|<1$ for all $l\in [1:N]$. This means that the stability constraint is reduced to
\begin{equation}\label{lambda_max}
|\lambda_{max}|<1,
\end{equation}
where $\lambda_{max}$ is the maximum among all the $N$ eigenvalues. To determine the eigenvalues of $J_s$ we diagonalize the matrix using Fourier matrix as a diagonalization matrix \cite{PM_Gade_1998} (note that, a circulant matrix can be diagonalized by using Fourier matrix of the same order). The diagonalized matrix $U_d$ is obtained by the operation
\begin{equation}\label{F_diagonal}
U_d=\frac{1}{N}(F^{-1}J_n F),
\end{equation}
where $F$ is the Fourier matrix of the order $N\times N$ whose elements are independent of the matrix to be diagonalized. The elements of the Fourier matrix are given by $F_{i,j}=\omega^{(i-1)(j-1)}$, where $i$ and $j$ are,  respectively, the row and column numbers of the respective elements in the matrix and $\omega$ is the $N^{th}$ root of unity. Therefore, the Fourier matrix takes the following: 
\begin{equation}
F=\left[
\begin{matrix}
\omega^{0} & \omega^{0} & \omega^{0} & \dots & \omega^{0}\\
\omega^{0} & \omega^{1} & \omega^{2} & \dots & \omega^{(N-1)}\\
\omega^{0} & \omega^{2} & \omega^{4} & \dots & \omega^{2(N-1)}\\
\omega^{0} & \omega^{3} & \omega^{6} & \dots & \omega^{3(N-1)}\\
\vdots & \vdots & \vdots & \vdots &\vdots &\\
\omega^{0} & \omega^{(N-1)} & \omega^{2(N-1)} & \dots & \omega^{(N-1)(N-1)}
\end{matrix}
\right],
\end{equation}
and the inverse of the above Fourier matrix can be written as
\begin{equation}
F^{-1}=\left[
\begin{matrix}
\omega^{0} & \omega^{0} & \omega^{0} & \dots & \omega^{0}\\
\omega^{0} & \omega^{-1} & \omega^{-2} & \dots & \omega^{-(N-1)}\\
\omega^{0} & \omega^{-2} & \omega^{-4} & \dots & \omega^{-2(N-1)}\\
\omega^{0} & \omega^{-3} & \omega^{-6} & \dots & \omega^{-3(N-1)}\\
\vdots & \vdots & \vdots & \vdots &\vdots &\\
\omega^{0} & \omega^{-(N-1)} & \omega^{-2(N-1)} & \dots & \omega^{-(N-1)(N-1)}
\end{matrix}
\right].
\end{equation}
Now, all the eigenvalues $\lambda_i$ are obtained by finding the diagonal elements of \eqref{F_diagonal}. The general analytical expression for the diagonal elements and hence the eigenvalues are found to be
\begin{widetext}
\begin{eqnarray}\label{eigenvals}
\lambda_d(l)&=& 1-K_1\xi\cos(\phi_s)+\frac{\epsilon K_1\xi}{R}\cos(\phi_s)\left[\cos(\frac{2\pi}{N}l)+\cos(\frac{2\pi}{N}2l)+\cos(\frac{2\pi}{N}3l)+\dots +\cos(\frac{2\pi}{N}Rl)\right] \nonumber \\ 
            &=& 1-K_1\xi\cos(\phi_s)+\frac{\epsilon K_1\xi}{R}\cos(\phi_s)\sum_{m=1}^R\cos(\frac{2\pi}{N}ml).
\end{eqnarray}
Therefore, the largest eigenvalue can be obtained by putting the values of corresponding $l$ and other parameters in the above expression. Then we use the condition \eqref{lambda_max} in \eqref{eigenvals} to determine the parameters for which the network attains a stable synchronized (phase-locked) state.
\end{widetext}
%%%%%%%%%%%%SECTION4:SIMULATION RESULTS%%%%%%%%%%%%%%%%%%%%%%%%%%%%%%
\section[Results: an one-dimensional network]{Results: one-dimensional network}
We carry out extensive numerical simulations on the system equation \eqref{system_eq_nonlocal_1d} taking a system size $N=256$ (we also verify our results for a large number of nodes and get the same qualitative results). In all the simulations we consider random initial conditions that are distributed uniformly in the range $[-\pi:\pi]$, which are the possible values of phases, $\{\phi\}$, of ZC1-DPLLs. 
We choose the value of the normalized frequency as $\xi=1.1$, which is the practical value used in the DPLL literature (generally named as frequency step input condition)~\cite{lin81,T_Banerjee_2014}. 
In all the simulations $5000$ initial time steps are discarded to avoid transients.

%%%%%%%%%%%%SUBSECTION%%%%%%%%%%%%%%%%%%%%%%%%%%%%%%
\subsection{Phase diagram}
\begin{figure}[!tbp]
  \begin{center}
    \leavevmode
    \ifpdf
      \includegraphics[height=3in]{}
    \else
      \includegraphics[width=.45\textwidth]{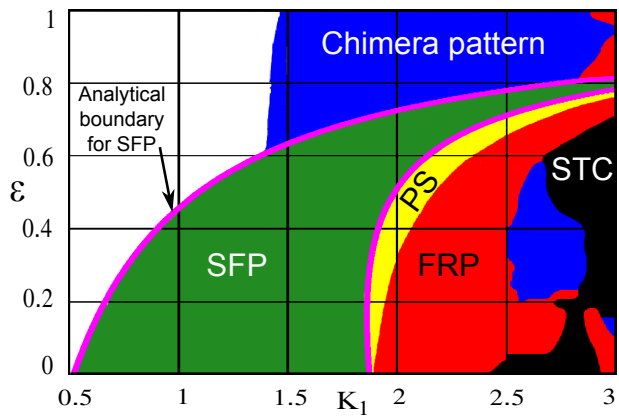}
    \fi
    \caption[Phase diagram showing different complex dynamics of the system]{(Color online) The phase diagram in the $K_1-\epsilon$ parameter space. The region of different dynamics are shown: SFP: Synchronized fixed point, FRP: Frozen random pattern, PS: Pattern selection, STC: spatiotemporal chaos, and chimera pattern for a network size $N=256$, coupling range $R=105$ and $\xi=1.1$. The thick (magenta) line around the SFP region gives the stability curve derived  analytically in \eqref{eigenvals} along with \eqref{lambda_max}: It matches with the numerical results. In the white region there is no distinguished recognizable regular spatiotemporal patterns rather it shows quasiperiodic and chaotic behaviors.}
\label{1dphasediagam}
  \end{center}
\end{figure}

Figure~\ref{1dphasediagam} presents the phase diagram showing several spatiotemporal dynamics in the $K_1-\epsilon$ parameter space for an exemplary coupling range $R=105$ (note that we choose the value of $R$ such that $1<R<\frac{N}{2}$). We map the parametric zone of occurrence of the following prominent dynamical states~\cite{kan1,kan2}: (i) synchronized fixed point (SFP) solutions, where all the DPLLs in the network are phase-locked or synchronized with each other to a common steady-state phase $\phi_s$ (given by \eqref{ss_1d_nonlocal_ch4}), (ii) pattern selection (PS), where the whole network is subdivided into several clusters of standing waves having well characterized wavelengths (to be discussed later), (iii) frozen random pattern (FRP), where cluster forms but unlike PS there exists no temporal variation of the clusters, (iv) spatiotemporal chaos (STC), where all the DPLLs show chaotic behavior both spatially and temporally, and (v) chimera state, i.e., the spatiotemporal coexistence of coherence and incoherence. 
%In the study of 
In the phase diagram we also plot the boundary corresponding to the SFP state obtained from the analytical expression derived in Eq.~\eqref{eigenvals} along with Eq.~\eqref{lambda_max}. It is apparent that the analytical and numerical results agree well with each other. 

%%%%%%%%%%%%SECTION:Chimera state%%%%%%%%%%%%%%%%%%%%%%%%%%%%%%
\subsection{Chimera state}\label{chimera_state}
\begin{figure}[!htbp]
  \begin{center}
    \leavevmode
    \ifpdf
      \includegraphics[height=3in]{}
    \else
      \includegraphics[width=.47\textwidth]{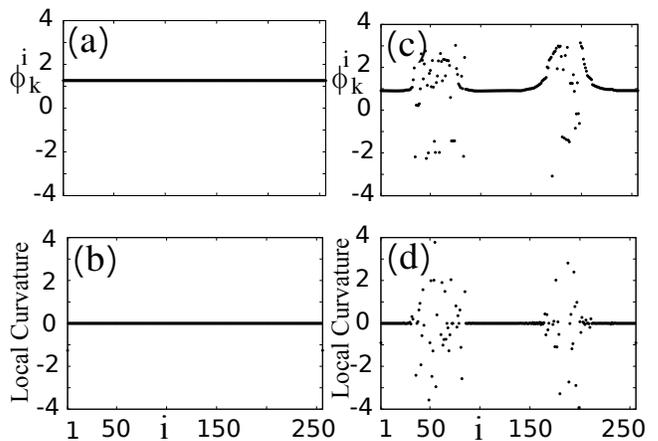}
    \fi
    \caption[SFP to chimera transition with coupling strength]{(Color online) The transition from synchronized fixed point (SFP) to chimera state with the increase in coupling strength $\epsilon$. (a),(b): SFP pattern with $\epsilon=0.7$. (c),(d): chimera pattern with $\epsilon=0.8$. Snapshot of $\phi{^i}{_k}$ (at $k=5100$) [(a) and (c)] of all the nodes, and the corresponding local curvature given by \eqref{1d_diff1} [(b) and (d)] demonstrate the coexistence of coherent and incoherent states in the chimera state and spatial synchrony in the SFP state. Other parameters: $K_1=2$ and $R=105$.}
\label{chimera11}
  \end{center}
\end{figure}

It is apparent from the phase diagram (Fig.~\ref{1dphasediagam}) that chimera pattern occurs in a broad parameter zone. In Fig.\ref{chimera11} we demonstrate the transition from the SFP state (Fig.~\ref{chimera11}(a)) to the chimera state (Fig.~\ref{chimera11}(c)) with the variation of coupling strength $\epsilon$ (at $K_1=2$). For $\epsilon=0.7$  SFP state occurs (Fig.\ref{chimera11}(a)), where  all the DPLLs are synchronized with a constant $\phi_s=1.258$ (as predicted analytically in \eqref{ss_1d_nonlocal_ch4}). With an increase in $\epsilon$ chimera state emerges: it is shown in Fig.~\ref{chimera11}(c) for $\epsilon=0.8$, where we can observe the coexistence of coherent and incoherent DPLLs.   

To ensure and characterize the emergence of the chimera state we use the concept of local curvature proposed recently by \citet{kemeth_2016}. It characterizes the spatial coherence among the neighboring sites~\cite{kemeth_2016}. The {\it local curvature} in one dimensional systems is defined as the second derivative of the local attributes (phase states) of the network with the space coordinate. 
In our present case it is defined as
%\begin{equation}
%\nabla \phi_k^{(i)}= \phi_k^{(i+1)}-\phi_k^{(i)}
%\end{equation}
%Then the second-order difference equation is such defined so that it becomes symmetric with respect to the center element $\phi_k^{(i)}$; therefore, it is written as 
\begin{eqnarray}\label{1d_diff1}
\nabla(\nabla \phi_k^{(i)})& = &\nabla \phi_k^{(i)}-\nabla \phi_k^{(i-1)},\nonumber \\
& = & \{\phi_k^{(i+1)}-\phi_k^{(i)}\}-\{\phi_k^{(i)}-\phi_k^{(i-1)}\}, \nonumber \\
& = & \phi_k^{(i+1)}-2\phi_k^{(i)}+\phi_k^{(i-1)}.
\end{eqnarray}
The local curvature for the SFP and the chimera states are shown in Fig.~\ref{chimera11}(b) and Fig.~\ref{chimera11}(d), respectively (at $K_1=2$ and $R=105$): for the coherent domain it becomes zero whereas in the incoherent domains it fluctuates randomly with non zero values. 
%\subsection{Bifurcation diagram}\label{cml_bifur_dia}

\begin{figure}[!htbp]
  \begin{center}
    \leavevmode
    \ifpdf
      \includegraphics[height=3in]{}
    \else
      \includegraphics[width=.4\textwidth]{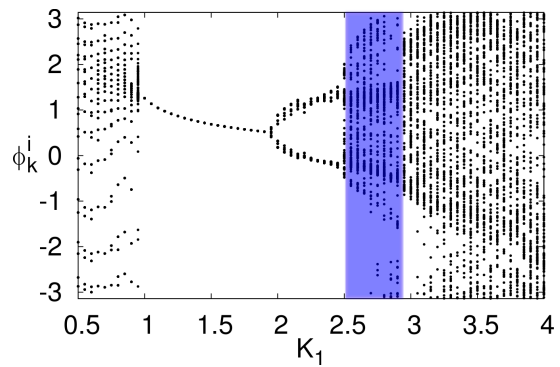}
    \fi
    \caption[The bifurcation diagram for the network coupled DPLLs]{(Color online) The bifurcation diagram for the whole network with the variation of gain $K_1$ ($R=105$ and $\epsilon=0.4$). The shaded zone indicates the parametric zone of occurrence of chimera.}
\label{cmlbifur}
  \end{center}
\end{figure}

To get an idea of the parameter zone of local dynamics where chimera occurs we plot the bifurcation diagram for the coupled network with the variation of gain parameter $K_1$ (shown in Fig.~\ref{cmlbifur} for $\epsilon=0.4$ and $R=105$). The shaded zone in Fig.~\ref{cmlbifur} indicates the zone of $K_1$ where chimera pattern emerges: it shows that  in the chimera state the local temporal dynamics is chaotic (we also verify it by computing Lyapunov exponents of the whole network--results not shown here). This zone is also instructive for the choice of $K_1$ in search of chimera in the 2D network of DPLLs (discussed in the next section). For other values of $R$ the qualitative structure of the bifurcation structure remains the same.

\begin{figure}[!htbp]
  \begin{center}
    \leavevmode
    \ifpdf
      \includegraphics[height=3in]{}
    \else
      \includegraphics[width=.48\textwidth]{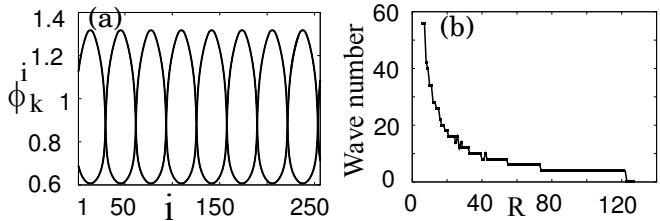}
    \fi
    \caption[Variation of wave length in pattern selection with the range of coupling]{(Color online) (a) The pattern selection (for $R=50$). The spatial domain is subdivided into eight domains. (b) The variation of wave number with $R$: the wave number decreases with increasing $R$. Other parameters: $K_1=2.75$, $\epsilon=0.73$.}
\label{ps1} 
  \end{center}
\end{figure}

%\subsection{Pattern selection and wave number}\label{waveno}
Further, we study how the coupling range $R$ affects the spatiotemporal patterns. For this we investigate the pattern selection (PS) dynamics and its dependence on the coupling range $R$: these are shown in Fig.~\ref{ps1}(a) and Fig.~\ref{ps1}(b), respectively. In the PS state the network is subdivided equally into several domains forming standing wave pattern and the dimension of each domain is equal to the wavelength of the formed standing wave. 
The number of wavelengths in the network is called wave numbers~\cite{kan1}. 
Figure~\ref{ps1}(a) shows the standing wave pattern for $R=50$ showing eight domains. As coupling range $R$ increases the number of standing waves decreases: this is shown in the Fig.~\ref{ps1}(b). This means that a larger $R$ prefers the SFP pattern (i.e., zero wave number), therefore, it leads to global phase-locked condition. It is quite expected as larger $R$ means near global coupling, which is conducive for the global synchrony. We verify that an intermediate range of $R$ supports chimera pattern to occur: lower $R$ leads to asynchrony and larger $R$ leads to global synchrony.  

%%%%
%%%%%%%%%%%%%%%%%%%%%%%%2D CML%%%%%%%%%%%%%%%%%%%%%%%%%%%%%%%%%%%%%%%%
\section{2D network: Coupling scheme and mathematical model}

\begin{figure}[!tbp]
  \begin{center}
    \leavevmode
    \ifpdf
      \includegraphics[height=3in]{}
    \else
      \includegraphics[width=.35\textwidth]{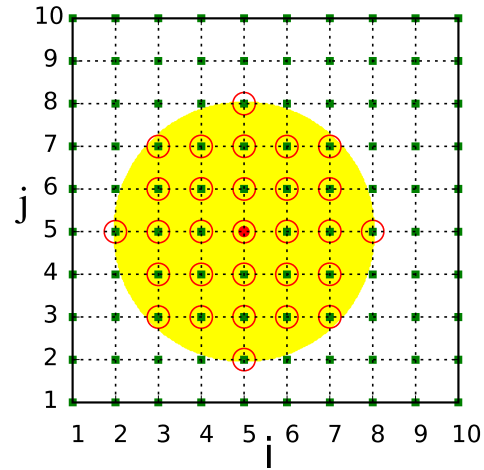}
    \fi
    \caption[Schematic diagram of the coupling scheme of a 2D network of coupled ZC1-DPLLs]{(Color online) The representative schematic diagram of the coupling scheme of a 2D network of nonlocally coupled ZC1-DPLLs. We show it for $N=10$ and coupling range $R=3$. Each  node of the mesh contains a ZC1-DPLL (shown with filled square) coupled with neighboring sites with periodic boundary condition (or toroidal boundary condition). The indices $i$ and $j$ represent the spatial co-ordinates in two mutually perpendicular direction of the 2D matrix. The coupling topology is shown for a particular site $i=j=5$ (depicted by the red solid circle). The node with $i=j=5$ is coupled with those nodes, which fall within the circle of influence (shown with (yellow) shading) with a fixed radial distance. The nodes that are coupled with the $(i,j)=(5,5))$-th site are encircled by red hollow circles.}
\label{2d_mesh_dpll}
  \end{center}
\end{figure}

We consider a two-dimensional (2D) network of coupled ZC1-DPLLs with orthogonal mesh of dimension $(N\times N)$. Figure~\ref{2d_mesh_dpll} shows the representative spatial layout of the 2D network with ZC1-DPLLs at each node of the mesh. We consider the toroidal boundary condition with nonlocal coupling scheme, i.e., a periodic boundary condition which is compatible with the 2D network structure. Each ZC1-DPLL at the nodes of the network is non-locally coupled with their neighbors that are located within the {\it circle of influence} whose area is determined by the radial distance from the node in consideration. Therefore, a ZC1-DPLL at the spatial position $(i,j)$ is coupled with those ZC1-DPLLs that are inside the circle of radius $R$ (in the unit of lattice constant of the mesh or lattice) centered at the position $(i,j)$. Here $R$ is called the coupling range: $R=1$ corresponds to the nearest neighbor coupling whereas $R=0$ refers to the condition of no coupling and the system is considered to be an ensemble of isolated $N^2$ number of ZC1-DPLLs.

%%%%%%%%%%%%%%%SECTION3:SYSTEM DESCRIPTION%%%%%%%%%%%%%%%%%%%%%%%%%%

We derive the system equation of the 2D network of nonlocally coupled ZC1-DPLLs by using the method used in Sec.~\ref{1dnet}: it is given by
\begin{eqnarray}\label{system_eq_nonlocal_2d}
%\begin{align}
\phi^{i,j}_{k+1} & = & \Lambda_0+\phi^{i,j}_k-\xi K_1 \sin\phi^{i,j}_k\nonumber \\
&& +\: \frac{\epsilon \xi K_1}{N_R-1}\sum_{m,n}\left[\sin(\phi^{m,n}_{k})-\sin(\phi^{i,j}_k)\right].
%\end{align}
\end{eqnarray}
%Eqn.~(\ref{syst_eq_iso_ch4}) and Eqn.~(\ref{system_eq_nonlocal_1d}) having the same form except the last term in Eqn.~(\ref{system_eq_nonlocal_1d}) due to the coupling in the network. 
The indices $i$ and $j$, indicating the spatial coordinates in the two mutually perpendicular directions, are bounded in the interval $i,j\in[1:N]$ having only integer values. The index $k$ represents temporal sequences of the evolutions. The indices $m$ and $n$ are used to include the effect of the coupling topology. The values of $m$ and $n$ for the $(i,j)$-th site is such that the coupling is realized within a circular region of radius $R$ centered at the $(i,j)$-th site; Therefore~\cite{Schmidt_2017},
\begin{equation}\label{def_m,n}
m,n\in[(m-i)^2+(n-j)^2\leq R^2],
\end{equation}
and 
\begin{equation}\label{def_N_R}
N_R=1+4\sum^{\infty}_{p=0}\left(\lfloor\frac{R^2}{4p+1}\rfloor-\lfloor\frac{R^2}{4p+3}\rfloor\right),
\end{equation}
where $N_R$ is the number of ZC1-DPLLs which are coupled to a particular node of the network when the coupling radius is $R$. Here $p$ can take only integer values. From Fig.~\ref{2d_mesh_dpll} the number of ZC1-DPLLs within the yellow shaded region is $N_R=29$ (including the red circle) with $R=3$. The symbol $\lfloor \cdot \rfloor$ represents the floor function that produces the largest integer less than or equal to its argument.
%Now like the isolated ZC1-DPLL, here also $\Lambda_0=2\pi(\xi-1)$ with $\xi=\frac{\omega}{\omega_0}$ is the normalized frequency of the incoming signal over the DCO nominal frequency and $K_1$ is the loop gain of each ZC1-DPLL. Here $\epsilon$ represent the coupling strength among the sites under coupled. The coupling range $R=0$ refers to the condition of no coupling among the DPLLs, then the system is considered to be an ensemble of isolated $N^2$ DPLLs. If $R=1$, then the nearest neighbour coupling is realised and it may be termed as local coupling.

\begin{figure}
\begin{center}
\includegraphics[width=.5\textwidth]{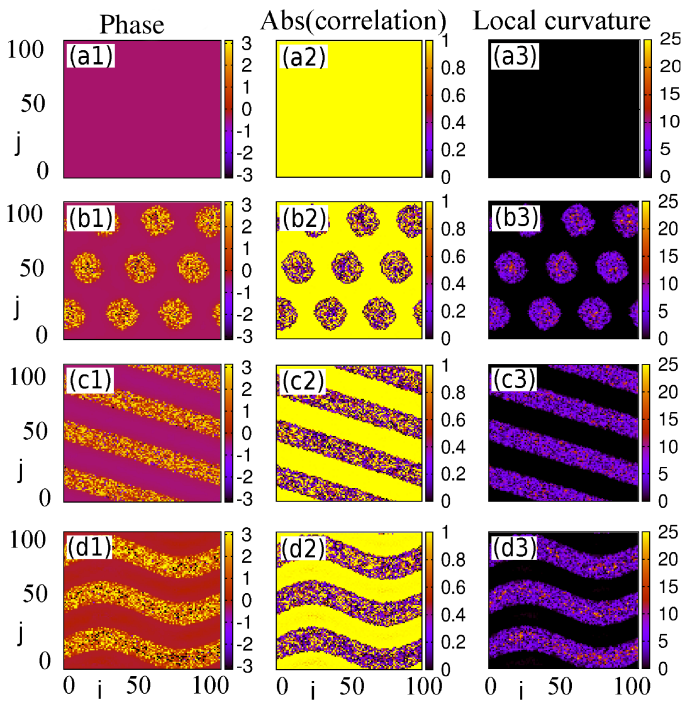}
\caption[change of chimera with coupling strength]{(Color online) (a1)--(d1) Plot of phase $\phi^{i,j}_{k}$ (taken at $k=5100$); (a2)--(d2) Respective plots of the absolute values of the local correlation coefficient and (a3)--(d3) Respective local curvatures. The figure shows the transition from SFP (a1-a3) for $\epsilon=0.2$ $\rightarrow$ grid chimera (b1-b3) for $\epsilon=0.3$ $\rightarrow$ strip chimera (c1-c3) for $\epsilon=0.4$ $\rightarrow$ wavy strip chimera (d1-d3) for $\epsilon=0.5$. Here $K_1=2.75$, $R=25$.}
\label{2dchimera} 
\end{center}
\end{figure}
%%%%%%%%%%%%%SECTION4:SIMULATION RESULTS%%%%%%%%%%%%%%%%%%%%%%%%%%%%%
\section{Results}
%An extensive simulation is performed to explore the dynamical nature due to non-local coupling in the proposed two-dimensional network of ZC1-DPLLs. 
Numerical simulations are carried out on the 2D-network of $(100\times 100)$ sites with toroidal boundary conditions. Similar to the 1D case here also we consider random initial conditions uniformly distributed in the range $[-\pi:\pi]$, and we take $\xi=1.1$. To discard the transients, all the plots are presented after discarding $k=5000$ iterations. 

\begin{figure}
\begin{center}
\includegraphics[width=.5\textwidth]{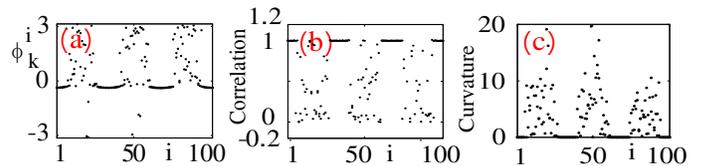}
\caption[Representation of local correlation coefficient and local curvature in two dimensional space]{(Color online) The cross sectional view of Figs.~\ref{2dchimera}(b1)--(b3) along $j=50$. (a) Snapshot of spatial variation of phase at $k=5100$, (b) absolute value of local correlation coefficient (c) the local curvature $D$. Parameter values are: $R=25$, $\epsilon=0.3$, $K_1=2.75$ and $L=1$.}
\label{cross2d} 
\end{center}
\end{figure}

Our simulation results reveal that there exist several 2D spatiotemporal patterns, which can be broadly classified  into two categories, namely chimera patterns and non-chimera patterns. The typically observed chimera patterns are grid chimera or spot chimera, strip chimera and wavy strip chimera. 
Figures~\ref{2dchimera}(a1)-(d1) illustrate the transition from the synchronized fixed point (SFP) pattern to several chimera patterns with the increase of coupling strength ($\epsilon$) for a fixed coupling range $R=25$ (guided by Fig.~\ref{cmlbifur} we take $K_1=2.75$, which lies within the blue shaded region). Figure~\ref{2dchimera}(a1) shows the SFP state for $\epsilon=0.2$: in this state all the DPLLs are phase-locked with each other. With an increase in $\epsilon$ {\it grid or spot chimera} pattern appears. This is shown in Fig.~\ref{2dchimera}(b1) for $\epsilon=0.3$: here incoherent spots emerge in the background of synchronized pattern (later we will see that diameter of these spots depend upon the coupling range $R$). Further increase of $\epsilon$ results in {\it strip chimera} pattern as shown in Fig.~\ref{2dchimera}(c1) for $\epsilon=0.4$: we see that here the incoherent domains exhibit strips of certain width (later we will see that width of the strip increases with increasing $R$). Finally, for a larger value of $\epsilon$ we get {\it wavy strip} chimera patterns (Figure~\ref{2dchimera}(d1) for $\epsilon=0.5$): here, interestingly, the incoherent domains (and thus the coherent domains) show a wave like pattern in spatial landscape. For a better understanding of the coherent-incoherent structure of the chimera pattern we show the cross section of Fig.~\ref{2dchimera}(b1) (i.e., the spot chimera) along $j=50$ in Fig.~\ref{cross2d}(a1): one can see three incoherent regions corresponding to the three incoherent spots of Fig.~\ref{2dchimera}(b1).  
 
%%%%%%%%%%%%%

%%%%%%%%%%%%%%%%%%%%%measures
To ensure the occurrence of chimera patterns we introduce two quantifiers that measures the spatial correlation among the different sites, namely the local correlation coefficient \cite{corr,T_Banerjee_2014} and the local curvature \cite{kemeth_2016}. Both these measures provide the degree of coherence among the nodes and therefore are capable of distinguishing coherent and incoherent domains. The local correlation coefficient is defined as \cite{corr},
%%%%%%%%%%%%%%%%%%
\begin{equation}\label{local_correlation}
%\begin{align}
(e_k^{i,j})_L=\frac{\sum_{m^\prime,n^\prime}\hat\phi^{i,j}_k\hat\phi^{m^\prime,n^\prime}_k}{\sum_{m^\prime,n^\prime}(\hat\phi^{m^\prime,n^\prime}_k)^2},
%\end{align}
\end{equation}
where $m^\prime,n^\prime$ refers to the sites that are within a circle of radius of $L$ centered at the site having index $(i,j)$. So,
\begin{equation}\label{L_def}
m^\prime,n^\prime\in(m^\prime-i)^2+(n^\prime-j)^2\leq L^2,
\end{equation}
$L$ is called the degree of correlation. For first order (local) correlation, $L=1$, the correlation is totally with the nearest neighbors; whereas for second order, $L=2$, includes the second nearest neighbors and so on. 
%Now as before,
%\begin{equation}\label{N_L_def}
%N_L=1+4\sum_{p=0}^\infty\left\{\left\lfloor\frac{L^2}{4p+1}\right\rfloor-\left\lfloor\frac{L^2}{4p+3}\right\rfloor\right\}.
%\end{equation}
The deviation of phases of the ZC1-DPLLs in the equation (\ref{local_correlation}) are follows:
\begin{equation}\label{phi_sum}
\hat\phi^{i,j}_k=\phi^{i,j}_k-<\phi^{i,j}_R>_k
\end{equation}
is the deviation of phase of the $(i,j)$-th site. Whereas
\begin{equation}\label{phi_sum_mn_prime}
\hat\phi^{m^\prime,n^\prime}_k=\phi^{m^\prime,n^\prime}_k-<\phi^{i,j}_R>_k
\end{equation}
is the deviation of phases of those sites that are within the range of order of correlation. For local correlation this range is a unit circle centered at $(i,j)$-th. Here $<\phi^{i,j}_R>_k$ is the average of the local phases up-to the coupling range $R$ of the system for the spatial index $(i,j)$ and is given by
\begin{equation}\label{phy_average}
<\phi^{i,j}_R>_k=\frac{1}{N_R}\sum_{m,n}\phi^{m,n}_k,
\end{equation}
where $m,n$ and $N_R$ are given in Eqns.~(\ref{def_m,n}) and (\ref{def_N_R}), respectively. The value of the local correlation co-efficient $\mbox{abs}\left((e_k^{i,j})_L\right)=1$ corresponds to the local synchronization, and the deviation of the value from unity means local incoherence.

Next, we use the measure of local curvature \cite{kemeth_2016}. In the 1D case, the general Laplacian is simply a second derivative as used in \eqref{1d_diff1}, however, for the 2D case the second order difference equation, i.e., the local curvature $D$ becomes
\begin{eqnarray}\label{2d_diff}
D&=&\nabla(\nabla \phi_k^{(i,j)}) = \nabla_i^2\phi_k^{(i,j)}-\nabla_j^2\phi_k^{(i,j)}, \nonumber \\
%&=& \phi_k^{(i+1,j)}-2\phi_k^{(i,j)}+\phi_k^{(i-1,j)}+\phi_k^{(i,j+1)}-2\phi_k^{(i,j)}+\phi_k^{(i,j-1)} \nonumber \\
&=& \phi_k^{(i+1,j)}+\phi_k^{(i-1,j)}+\phi_k^{(i,j+1)}+\phi_k^{(i,j-1)},\nonumber \\
&& -\: 4\phi_k^{(i,j)},
\end{eqnarray}
which is symmetric with respect to the center element $\phi_k^{(i,j)}$. $D=0$ corresponds to the coherent and $D\ne 0$ refers to the incoherent domains. These two measures are shown in Figs.~\ref{2dchimera} to support our results: the middle column [Figs.~\ref{2dchimera}(a2)--(d2)] represents the absolute value of the local correlation coefficient and the right most column [Figs.~\ref{2dchimera}(a3)--(d3)] shows the local curvature $D$. In Figs.~\ref{2dchimera} (a2)--(d2) the coherent nodes have the (absolute) local correlation coefficient equal to unity [shown in light gray (yellow) shading], whereas the incoherent nodes differ from unity [see the color map in Figs.~\ref{2dchimera}(a2)--(d2)]. Equivalently, in Figs.~\ref{2dchimera}(a3)--(d3) the local curvature $D$ for the coherent nodes is zero and that for incoherent domains are nonzero.  Fig.~\ref{cross2d}(b) and  Fig.~\ref{cross2d}(c) represent this scenario clearly along the cross section of Fig.~\ref{2dchimera}(b1) at $j=50$ for the spot chimera: in the incoherent zones the local correlation coefficient attains a value less than unity whereas the local curvature $D$ jumps from its zero value to larger nonzero values. 

\begin{figure}
\begin{center}
\includegraphics[width=.4\textwidth]{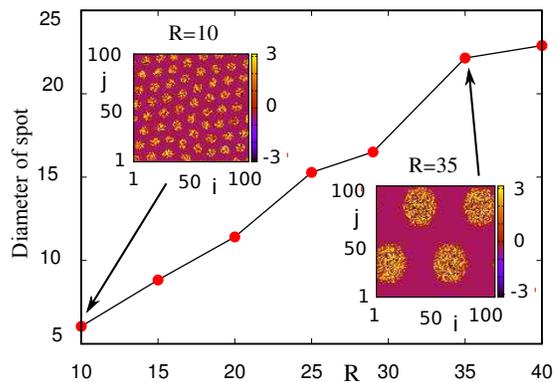}
\caption[Variation of diameter of circular chaotic spots with the range of coupling $R$]{(Color online) Variation of diameter of circular incoherent spots with the range of coupling $R$ for fixed $\epsilon=0.3$ and $K_1=2.75$.}
\label{diameter_grid}
\end{center} 
\end{figure}

\begin{figure}[!btp]
\begin{center}
\includegraphics[width=.4\textwidth]{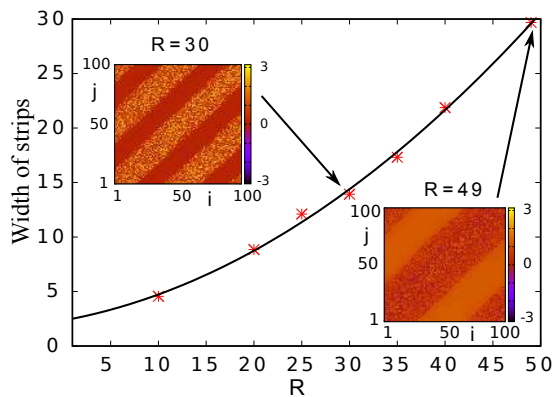}
\caption[Variation of chaotic strips width with the range of coupling]{(Color online) Variation of incoherent strip width with the coupling range $R$ (shown in points) for fixed $\epsilon=0.6$ and $K_1=2.75$. The curve is fitted with the polynomial $ax^2+bx+c$, where $a=0.008128$, $b=0.1579$ and $c=2.338$ (solid line).}
\label{wavelength_strip}
\end{center}
\end{figure}
Next, we investigate the dimension of incoherent regions in the case of grid (spot) and strip chimeras. We find that in the case of grid chimera, the dimension of the incoherent spot increases with coupling range $R$. 
The diameter of a grid is measured as the fractional percentage with respect to the dimension of 2D space lattice.
Therefore, the number of incoherent spots decreases with an increase in $R$. This is shown in Fig.~\ref{diameter_grid} ($\epsilon=0.3$, $K_1=2.75$): two representative patterns of the grid chimera for $R=10$ and $R=35$ are also shown in the inset showing the increase in spot size with increasing $R$.
In the case of the strip chimera the width of the incoherent strips (therefore, also, that of the coherent strips) increases with increasing $R$. 
The width of the strips are measured as the fractional percentage with respect to the dimension of diagonal length of the total 2D space lattice.
This scenario is shown in Fig.~\ref{wavelength_strip} ($\epsilon=0.6$, $K_1=2.75$): it can be seen that the width of the incoherent strips increases monotonically obeying a quadratic polynomial function (obtained by numerically fitting the data). Also shown are two representative strip chimera patterns for $R=30$ and $R=49$. Interestingly, we observe that the dimension of spots or the width of the strips are independent of the variation of value of coupling parameter $\epsilon$ and the gain constant $K_1$ (not shown here).

%%%%%%%%%%%%%%%%%%%%%%%

\begin{figure}[!thbp]
  \begin{center}
    \leavevmode
    \ifpdf
      \includegraphics[height=3in]{}
    \else
      \includegraphics[width=.48\textwidth]{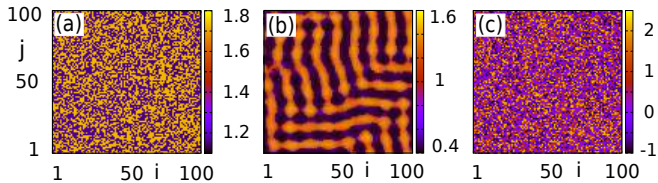}
    \fi
    \caption[Sptiotemporal patterns that are other than the chimera pattern]{(Color online) Non-chimera patterns: (a) spatially irregular period-two pattern ($R=10$, $\epsilon=0.10$, $K_1=2.75$), (b) spatial multiple periodic solutions ($R=10$, $\epsilon=0.70$, $K_1=2.70$), and (c) spatial chaos ($R=20$, $\epsilon=0.10$, $K_1=2.75$).}
\label{non-chimera}
  \end{center}
\end{figure}

\begin{figure}[!h]
\begin{center}
\includegraphics[width=.45\textwidth]{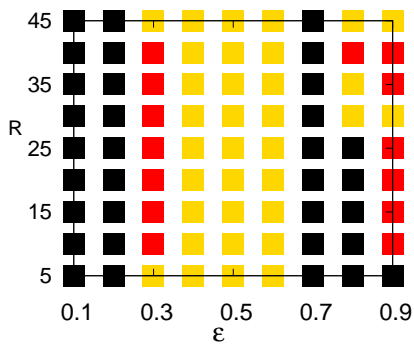}
\caption[Phase diagram in $R-\epsilon$ parameter space representing the different spatial patterns]{(Color online) The phase diagram representing different spatial patterns in the $R-\epsilon$ parameter space (for a fixed $K_1=2.75$). Color code: Black represents non-chimera patterns, Red represents grid chimera pattern, and  Yellow is for strip (linear and wavy) chimera patterns.}
\label{2Dphase}
\end{center}
\end{figure}

As discussed in the beginning of this section, apart from chimera patterns we also observed other non-chimera spatiotemporal patterns: these are spatially irregular period-two [Fig.~\ref{non-chimera}(a)], spatially multiple periodic [Fig.~\ref{non-chimera}(b)] and spatial chaos [Fig.~\ref{non-chimera}(c)] patterns for different choices of $R$, $\epsilon$ and $K_1$. These regular non-chimera patterns are demonstrated in the Fig.~\ref{non-chimera} for some exemplary parameters. The phase diagram of the parametric zone of occurrence of chimera patterns with respect to the non-chimera pattern is shown in Fig.~\ref{2Dphase} in the $\epsilon-R$ space at a fixed value of $K_1$. The black squares represent non-chimera states whereas gray (red) and light gray (yellow) squares represent grid (spot) and strip (linear and wavy) chimera patterns, respectively. It is noteworthy that at much lower value of $\epsilon$ chimera patterns do not appear. 


\section{Conclusion}
In this paper we have studied 1D and 2D networks of DPLLs with nonlocal coupling and shown that networks of DPLLs show several spatiotemporal instability including chimera states. For the 1D network we have carried out a linear stability analysis to identify the zone of stable phase-locked (synchronized) state in which DPLLs are desired to operate in any real applications. For the 2D network we have demonstrated that outside the zone of globally synchronized state several chimera patterns, such as spot and strip chimeras emerge. With suitable measures like local curvature and local correlation coefficient we have ensured and characterized the occurrence of the chimera patterns. Note that, in this paper we have established that a real engineering system even under its {\it normal coupling condition} may show chimera patterns. 
%Since the system and the coupling both are practically realizable therefore we believe that chimera patterns can be observed experimentally in this system.

From technical point of view, our study clearly set a road map for the designers to choose the coupling and system parameters such that a network of ZC1-DPLLs operates in the normal phase-locked condition. Further, in the context of control, this study reveals that only bifurcation  and chaos control in isolated DPLLs are not sufficient to ensure their normal phase-locked condition, rather in a network of DPLLs effective control schemes have to be implemented to tame chimera and other spatiotemporal states.

Interestingly, the experimental chimera pattern was first observed in a coupled map lattice (CML) system by \citet{raj}: typically, realization of a CML system in experiment is itself a challenging task. In this paper we also propose a system architecture which naturally describes a CML system and shows chimeras. Since DPLLs can be implemented using inexpensive FPGA and DSP boards \cite{dpll_book,dan2,dan3}, therefore, we believe that chimeras can also be observed in this network experimentally. Moreover, the occurrence of chimeras in DPLLs may arouse the attention of researchers to explore the possibility of exploiting chimeras in electronic communication and signal processing systems.  
\begin{acknowledgments}
Authors convey their sincere thanks to Prof. B. C. Sarkar for introducing them to the field of phaselock techniques. T.B. acknowledges the fruitful discussion with Prof. Eckehard Sch\"oll at TU-Berlin during his visit supported by DFG in the framework of SFB 910. 
\end{acknowledgments}


%\bibliography{chimera}
\end{document}